\documentclass[conference]{IEEEtran}
\IEEEoverridecommandlockouts

\usepackage{cite}
\usepackage{amsmath,amssymb,amsfonts}
\usepackage{dsfont}
\usepackage{algorithm}
\usepackage{algpseudocode} 
\usepackage{subfig}
\usepackage{url}
\usepackage{graphicx}
\usepackage{textcomp}
\usepackage{xcolor}
\def\BibTeX{{\rm B\kern-.05em{\sc i\kern-.025em b}\kern-.08em
    T\kern-.1667em\lower.7ex\hbox{E}\kern-.125emX}}
\begin{document}

\title{Discovering Communities in Continuous-Time Temporal Networks by Optimizing L-Modularity} 


\newtheorem{theorem}{Property}
\newtheorem{definition}{Definition}


\author{\IEEEauthorblockN{Victor Brabant}
\IEEEauthorblockA{\textit{UCBL, CNRS, INSA Lyon, LIRIS,} \\
\textit{UMR5205, F-69622}\\
Villeurbanne, France \\
victorbrabant@liris.cnrs.fr}
\and
\IEEEauthorblockN{Angela Bonifati}
\IEEEauthorblockA{\textit{UCBL \& IUF, CNRS, LIRIS } \\
\textit{UMR5205, F-69622}\\
Villeurbanne, France \\
angela.bonifati@univ-lyon1.fr}
\and
\IEEEauthorblockN{Rémy Cazabet}
\IEEEauthorblockA{\textit{UCBL, CNRS, INSA Lyon, LIRIS,} \\
\textit{UMR5205, F-69622}\\
Villeurbanne, France \\
remy.cazabet@univ-lyon1.fr}
}

\maketitle

\begin{abstract}
Community detection is a fundamental problem in network analysis, with many applications in various fields. 
Extending community detection to the temporal setting with exact temporal accuracy, as required by real-world dynamic data, necessitates methods specifically adapted to the temporal nature of interactions.
We introduce LAGO, a novel method for uncovering dynamic communities by greedy optimization of Longitudinal Modularity, a specific adaptation of Modularity for continuous-time networks.
Unlike prior approaches that rely on time discretization or assume rigid community evolution, LAGO captures the precise moments when nodes enter and exit communities. 
We evaluate LAGO on synthetic benchmarks and real-world datasets, demonstrating its ability to efficiently uncover temporally and topologically coherent communities.
\end{abstract}

\begin{IEEEkeywords}
temporal networks, community detection, dynamic communities, link stream, modularity.
\end{IEEEkeywords}

\section{Introduction}

Community detection is an important task in network analysis. It is used to uncover structural patterns and to reduce the complexity of large-scale graphs. Community detection has applications in many domains where systems can be modeled as networks, such as social science, economics, and biology.
In the static setting, leading approaches such as Louvain~\cite{Blondel_2008}, Infomap~\cite{Rosvall_2009}, or Leiden~\cite{Traag_2019} typically rely on defining an objective function and optimizing it using greedy algorithms. 
This approach offers two main advantages: it produces communities that are meaningful according to a well-defined quality measure, and it scales efficiently to large graphs due to the computational simplicity of greedy methods.

Real-world data often involves temporal dynamics, where interactions occur at specific timestamps. Extending community detection to such temporal networks requires methods that are specifically adapted to the temporal structure of interactions.

We argue that effective dynamic community detection should satisfy two key requirements: 
(i) flexibility in nodes joining and leaving a community, provided that the community maintains temporal coherence; and (ii) the ability to operate at the finest temporal resolution, directly on link streams \cite{latapy2018stream}, i.e., temporal networks of interactions without aggregation.
Although many methods for dynamic community detection have been proposed\cite{rossetti2018community}, none fully satisfy both criteria. Some approaches address (i) but require representing the temporal network as a multislice structure \cite{Mucha_2010}\cite{ infomap_multilayer}, which imposes a time scale and requires the temporal aggregation of high-resolution or sparse interaction data, therefore failing to meet criteria (ii). Other methods address (ii) by operating directly on link streams, but are designed to uncover non-evolving communities \cite{matias_2015}, or allow only a limited number of community transitions based on a predefined time granularity \cite{bovet2022flow}.

To address this limitation, Longitudinal Modularity \cite{brabant2025longitudinal} was recently introduced as the first quality function specifically designed to evaluate dynamic communities on link streams. Communities can be freely expressed as sets of time intervals of nodes.

However, no algorithm has yet been proposed to optimize this function. In this article, we bridge this gap by introducing LAGO (Longitudinal Agglomerative Greedy Optimization). It is the first method for detecting dynamic communities on link streams through greedy optimization of Longitudinal Modularity. LAGO allows uncovering the exact moments when nodes enter and leave communities, without need for aggregation or defining a temporal resolution.
As the first work proposing greedy exploration of link streams, our study provides an initial overview of the performance trends associated with different greedy strategies for constructing dynamic communities.
Moreover, LAGO is not limited to Longitudinal Modularity; it can be used to optimize any quality function that satisfies similar structural properties, making it possible to uncover dynamic communities tailored to specific criteria.

We begin by discussing the existing methods that form the foundation of our approach in Section \ref{sec:relwork}, followed by formal definitions and the necessary tools for working with continuous-time networks in Section \ref{sec:ctn}. LAGO is introduced in Section \ref{sec:lago} and its performance is evaluated and discussed in Section \ref{sec:experiments}.

\section{Related Work}\label{sec:relwork}

This section provides a brief overview of the key existing methods from which we take inspiration for our approach.
The most widely adopted methods for community detection in static networks typically follow a two-stage approach: (i) defining a quality function to evaluate the coherence of community structures; and (ii) employing a greedy algorithm to optimize this function.
We focus on three methods—each designed to optimize a different quality function. Despite their differing objectives, these methods share a common algorithmic structure, making them interchangeable in principle for optimizing the corresponding quality criterion. LAGO is designed by proposing adaptations of the algorithmic strategies of these methods to the continuous temporal setting.

\subsection{Louvain}

The Louvain method \cite{Blondel_2008} proposes an iterative and agglomerative algorithm optimizing Modularity \cite{newman2004finding} for detecting communities in static networks.

The algorithm proceeds through the iterative repetition of two phases until no more improvement in Modularity is possible. In the first phase, each node is initially assigned to its own community. Then, in a random order, each node is considered for reassignment to the community of one of its neighbors, selecting the move that maximizes the gain in Modularity. If any node is moved during an iteration, another full pass over all nodes is performed. This process is repeated until no reassignment leads to further improvement.
In the second phase, a new weighted graph is built, where nodes represent the communities identified in the first phase, and edge weights correspond to the sum of the inter-community link weights from the original graph. The first phase is then reapplied to this aggregated graph. These two phases are repeated alternately until Modularity can no longer be increased.

The method leverages the property that a partition defined on the aggregated graph preserves the same modularity value as the corresponding partition on the original graph. At each level of aggregation, the algorithm updates modularity values using only local information at that level. This design enables fast computation and allows the Louvain method to scale effectively to very large networks.

In the following, references to Louvain refer specifically to the algorithmic procedure itself, independent of the particular quality function being optimized.

\subsection{Infomap}\label{subsec:infomap}

Infomap is based on the optimization of the Map Equation \cite{Rosvall_2009}. It shares properties with modularity that allow for similar optimization strategy. Indeed, Infomap begins with a Louvain procedure --optimizing the Map Equation-- to produce an initial community partition. It then proceeds to a refinement phase consisting of two iterative steps: (i) single-node movements, in which nodes from the original graph are reassigned between the communities identified at the highest level of aggregation; and (ii) submodule movements, where the Louvain is applied within each module to identify submodules that are then allowed to move between higher-level modules. This two-phase refinement strategy enables Infomap to improve the community structure beyond the initial coarse partition. The two steps are applied repeatedly until the Map Equation cannot be further optimized.

\subsection{Leiden}

The Leiden \cite{Traag_2019} method was proposed to address limitations of the Louvain algorithm, such as the formation of disconnected subgraphs inside communities and the absence of guaranteed improvement at each optimization loop. These limitations persist regardless of the quality function optimized, including both the Constant Potts Model \cite{PhysRevE.84.016114}--which Leiden explicitly aims to optimize--and Modularity. Leiden introduces two key improvements. First, it integrates a refinement step after the first phase of each Louvain iteration to ensure that all communities remain internally connected. Unlike Infomap, where refinement is applied as a separate post-processing step, this connectivity refinement is embedded directly into the Louvain algorithm. Second, Leiden modifies the exploration strategy by maintaining a dynamic set of candidate nodes for reassignment. At each iteration, a node is selected from this set and evaluated for possible reassignment to a neighboring community. If the node is moved, its neighbors are added to the candidate set. This strategy improves algorithm speed. In the end, those improvements help avoid poor local optima, making Leiden both more accurate and more scalable than Louvain.

\section{Continuous-Time Networks}\label{sec:ctn}

Temporal networks are often modeled as sequences of snapshots, leading to the following drawbacks: (i) it is not possible to model fast time-evolving networks, unless (ii) an arbitrary aggregation window is chosen, leading to a loss of information.
Instead, modeling continuous time networks as link streams enables preserving the original time accuracy of the real phenomena observed. However, it requires designing new methods and optimization strategies, and redefining the notion of communities in this context.

\subsection{Link Streams}

\begin{definition}
A \textbf{link stream}  \cite{latapy2018stream} $\mathcal{L}$ is defined by a triplet $(T, V, E)$ where $T \subset \mathbb{R}$ is a time interval, $V$ a finite set of $N \in \mathbb{N}$ nodes, and $E = \{(uv, t) \in V^2 \times T\}$ a finite set of interactions.
\end{definition}
In this paper, we only consider the case where interactions are instantaneous, undirected, and unweighted. Also, we consider $T$ to be discrete. It does not affect our contribution since for real-world temporal interaction data, there is a natural timestep for a discretization that preserves exact timing which is the greatest common divisor of all the durations between successive timestamps of edges.

Let $uv_t = 1$ if $(uv, t) \in E$, else $0$. We define:
\begin{align}
    L_{uv, T'} = \sum_{t \in T'} uv_t
\end{align}
as the number of interactions between nodes $u$ and $v$ over a subset of time $T' \subset T$, and $L_{uv} = L_{uv, T}$ denotes the total number of interactions between nodes $u$ and $v$. Similar to static graphs, $k_u$ denotes the degree of node $u$ and $m$ the total number of interactions in the link stream:
\begin{align}
    k_u = \sum_{v \in V} L_{uv} \text{; }
    m = \sum_{u \in V} k_u / 2 = |E|
\end{align}

\subsection{Temporal Communities}

In this article, we follow the definition of dynamic communities introduced by Brabant et al.\cite{brabant2025longitudinal}:
\begin{definition}
A dynamic community structure over a link stream is defined as a collection of non-empty and mutually exclusive communities composed of sets of node-time pairs $\{(u_1 t_1), (u_1 t_2), ..., (u_2 t_3), (u_2 t_4), ...\}$.
\end{definition}
This formulation respects the non-overlapping nature of community structures while allowing two key temporal behaviors: (i) community membership may evolve over time, and (ii) nodes may remain outside of any community during periods of inactivity.
The latter is especially important in practice, as real-world networks often contain inactive periods—for instance, during nights in high-resolution datasets, or for nodes that have silently exited the system. The definition enables perfect precision of the time interval for nodes membership in communities.

For a community $C$ and a node $u$, we define:
\begin{align}
    T_{u \in C} = \{t \in T \text{ s.t. } ut \in C\}
\end{align}
as the set of time instants during which node $u$ belongs to community $C$,
\begin{align}
    T_{C} = \bigcup_{u \in V} T_{u \in C}
\end{align}
as the existence time of community $C$, and
\begin{align}
    L_{uv \in C} = L_{uv, T_{u \in C} \cap T_{v \in C}}
\end{align}
as the number of time edges between nodes $u$ and $v$ inside community $C$.

\begin{figure}[h]
    \centering
    \includegraphics[width=.5\textwidth]{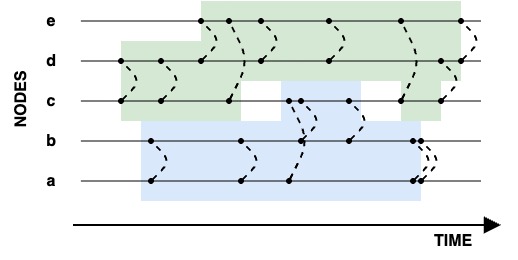}
\caption{Representation of a link stream with two dynamic communities. Nodes are represented in ordinates, and interactions between them occur through time.}
\label{fig:linkstream_example}
\end{figure}

\subsection{Longitudinal Modularity}

To evaluate the quality of dynamic communities on link streams, Brabant et al. introduced\cite{brabant2025longitudinal} the Longitudinal Modularity (L-Modularity), an adaptation of the modularity. Similar to its static counterpart, this function compares the observed fraction of interactions within communities to the expected fraction under a \textit{longitudinal random null model}. The greater this difference, the higher the L-Modularity. 

To account for the temporal dimension, two variants of L-Modularity were introduced: one based on the \textbf{Joint-Membership Expectation (JM)} and the other on the \textbf{Mean-Membership Expectation (MM)}.

JM \eqref{eq:le:jm} expects the overall structure of the community to be stationary or nearly stationary, with most of the nodes remaining in the same community throughout its duration. 
\begin{align}\label{eq:le:jm}
     \mathbb{E}_{JM}\left[L_{uv \in C} \right] = \dfrac{k_{u} k_{v}}{2m} \dfrac{|T_C|}{|T|} \mathds{1}_{|T_{u\in C}||T_{v\in C}| > 0}
\end{align}

MM \eqref{eq:le:mm} is more permissive, favoring more changes in node affiliations during the lifetime of a community.
\begin{align} \label{eq:le:mm}
     \mathbb{E}_{MM}\left[L_{uv \in C} \right] = \frac{k_{u} k_{v}}{2m} \frac{\sqrt{|T_{u \in C}| |T_{v \in C}|}}{|T|} 
\end{align}
To quantify the temporal smoothness of dynamic communities, L-Modularity includes a regularization term based on the sum of the \textbf{Community Switch Counts} (CSC) over all nodes, i.e. $\sum \eta_u$, where $\eta_{u}$ denotes the CSC of node $u$. The CSC quantifies how frequently a node changes its community membership over time. More precisely, $\eta_u$ is the number of communities visited by node $u$, minus one. The regularization term is weighted with a parameter $\omega \geq 0$ acting as a time resolution parameter. If $\omega = 0$, L-Modularity promotes communities with instantaneous time of existence, whereas higher values promote more stability in nodes affiliation to communities over time.

\begin{definition}
    With the previous notations, the \textbf{Longitudinal Modularity} of a dynamic community set $\mathcal{C}$ on a link stream $L$ regarding a time parameter $\omega > 0$ is given by:
    \begin{align}
         Q_{\star } (L, \mathcal{C}, \omega) = 
         & \frac{1}{2m} \sum_{C \in \mathcal{C}} \sum_{u,v \in V^2} \left[ L_{uv \in C} - \mathbb{E}_{\star}[L_{uv \in C}]\right] \nonumber \\
         & - \frac{\omega}{2m} \sum_{u \in V} \eta_{u}(\mathcal{C})
    \end{align}
    where $\star = JM \text{ or } MM$. 
\end{definition}
The optimization methods we propose are compatible with both versions of L-Modularity. 

The term \textit{time modules} refers to sub link streams that exhibit both temporal and topological coherence with respect to their contribution to the L-Modularity score. Strictly speaking, the optimization of L-Modularity yields time modules, which are subsequently interpreted as dynamic communities

\subsection{Trimmed Communities Property} \label{subsec:trimmed_communities}

\begin{figure}[h]
\centering
\subfloat[Link stream with dynamic communities in green and blue]{\includegraphics[width=.48\linewidth]{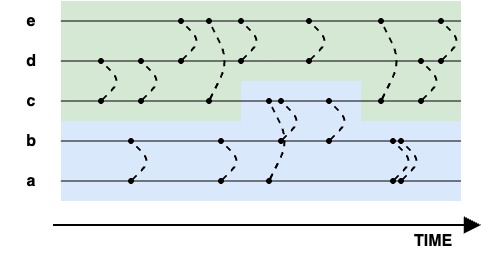}\label{figure:tcp:a}}
\subfloat[Trimmed version: inactive time nodes at boundaries are unaffiliated]{\includegraphics[width=.48\linewidth]{figures/linkstream_example.jpg}
\label{fig_second_case}\label{figure:tcp:b}}
\caption{Illustration of trimmed communities. L-Modularity assigns a higher value to the trimmed community structure that starts and ends on \textit{active time nodes} represented with black dots.} \label{figure:tcp}
\end{figure}

Building upon the previously introduced concepts, we define the \textit{trimmed communities property} which underpins the proposed algorithm. In essence, the property states that the L-Modularity score decreases when a community includes inactive nodes at its temporal boundaries. Consequently, communities can be trimmed to the earliest and latest active times of each of their nodes, as shown in Fig.~\ref{figure:tcp}.
In other words, L-Modularity favors community memberships that are strictly anchored to periods of observed activity and penalizes artificial extensions of membership into inactive periods.

We introduce $A$ the set of \textit{active time nodes} consisting of time nodes that interact:
\begin{align}\label{eq:atn}
    A = \{ut \in V \times T \; | \; \exists \; v \in V \text{, } (uv, t) \in E \} 
\end{align}
which allows to consider $T_{u \in C \cap A}$ the set of time instants where node $u$ is active inside a community $C$.

\begin{definition} \label{def:tnctrim}
    The \textit{trimmed existence time of node $u$ in community $C$} is
\begin{align} 
\lfloor T_{u \in C} \rfloor = T_{u \in C} \bigcap \left[ \bigwedge T_{u \in C \cap A}, \bigvee T_{u \in C \cap A} \right]
\end{align}
where $\bigwedge$ [resp. $\bigvee$] denotes the minimum [resp. maximum] element of a set. In other words, $\lfloor T_{u \in C} \rfloor $ is the smallest subset of $T_{u \in C}$ of all time instants between the first and last time instant where node $u$ is active within $C$. 
\begin{definition} \label{def:tctrim}
The \textit{trimmed existence time of community $C$} is
\begin{align}\label{def:tctrim}
    \lfloor T_C \rfloor  = \bigcup_u \lfloor T_{u \in C} \rfloor
\end{align}
\end{definition}

\begin{definition} \label{def:ctrim}
The \textit{trimmed community} $\lfloor C \rfloor$ is
\begin{align}
    \lfloor C \rfloor = \bigcup_{u \in V} u \times \lfloor T_{u \in C} \rfloor 
\end{align}
\end{definition}
By extension we note the set of trimmed communities as $\lfloor \mathcal{C} \rfloor = \{ \lfloor C \rfloor\}_{C \in \mathcal{C}}$.
Note that $T_{u \in \lfloor C \rfloor} = \lfloor T_{u \in C} \rfloor$ and $T_{\lfloor C \rfloor} = \lfloor T_C \rfloor$.
\end{definition}

\begin{theorem}[Trimmed Communities Property]\label{property:tcp}
Let $L$ be a link stream, $\mathcal{C}$ a dynamic community structure on it, and $\lfloor \mathcal{C} \rfloor$ its trimmed version. Then
    \begin{align}\label{eq:tcom}
    Q_{\star} (L, \lfloor \mathcal{C} \rfloor, \omega) \ge Q_{\star} (L, \mathcal{C}, \omega)
\end{align}
where $Q_{\star}$ denotes L-Modularity with $\star = JM \text{ or } MM$, and $\omega > 0$ is the term for time continuity constraint.
\end{theorem}

\begin{IEEEproof}
(i) The number of interactions does not change whether the time community is trimmed or not, i.e., $L_{uv \in \lfloor C \rfloor} =  L_{uv \in C}$; (ii) $\eta_u$ is not impacted by the durations of communities, only by their numbers, which is not impacted by trimming; (iii) longitudinal expectations have lower values in the trimmed version. Indeed, by Definition \ref{def:tnctrim}, 
\begin{align}\label{eq:ineqtuc}
    |\lfloor T_{u \in C} \rfloor| \leq | T_{u \in C} |
\end{align}
and then,
\begin{align}
    \mathbb{E}_{MM}[L_{uv \in \lfloor C \rfloor}] \leq \mathbb{E}_{MM}[L_{uv \in C}]
\end{align}
On the other hand, Definition \ref{def:tctrim} and \eqref{eq:ineqtuc} lead to $|\lfloor T_{C} \rfloor| \leq  | T_{C} |$, meaning that,
\begin{align}
    \mathbb{E}_{JM}[L_{uv \in \lfloor C \rfloor}] \leq \mathbb{E}_{JM}[L_{uv \in C}]
\end{align}
\end{IEEEproof}

This property implies that, with respect to L-Modularity, it is sufficient to consider only the set of active time nodes \eqref{eq:atn} when focusing on dynamic communities. Including other time nodes will never increase the L-Modularity value. Consequently, this reduces the amount of information that must be processed by an optimization algorithm, potentially improving computational efficiency.

\section{Proposed Algorithm}\label{sec:lago}

This section details our L-modularity optimization method, LAGO (Longitudinal Agglomerative Greedy Optimization), which is the core contribution of the paper. LAGO follows a greedy optimization approach, inspired by existing methods developed for static networks (see Section \ref{sec:relwork}). While preserving the general principles of these methods, we adapt the strategy to the specifics of L-Modularity and link streams.
 
We propose 14 LAGO variants, each corresponding to a different combinations of the strategies for exploring the space of possible time modules that we introduce in this section. A summary of the variants is provided in Table \ref{tab1}.

\subsection{Initialization step} 

Agglomerative methods for community detection on static networks are typically initialized with each node belonging to its own community, i.e., the finest possible partition of the network. 

When temporal networks are modeled as ordered multi-slice networks---as in the multi-slice modularity\cite{Mucha_2010} and Infomap adaptation to the multi-slice setting \cite{infomap_multilayer}--- the finest scale elements are temporal nodes, i.e., a pair $ut$ corresponding to a node $u$ at a time $t$. This modeling choice results in an initialization where each $ut$ instance is placed in its own community. Note that this is true even if the node is present but inactive.

A possible extension of this approach to link streams would consist of defining an arbitrary time step, and split the continuous lifetime of nodes accordingly, to create basic node units. One could use an arbitrary time step duration, or use the finest possible one, i.e., the greatest common divisor $d$ of all the durations between successive timestamps of edges, ensuring no loss of temporal information. However, this leads to $|V|\frac{|T|}{d}$ elements to consider, disregarding the actual temporal activity of the network, which is often significantly sparser, thereby introducing unnecessary computational overhead.

Instead, in LAGO, we propose to focus solely on the set of active time nodes, the time nodes that interact, as defined in \ref{eq:atn}.
According to the trimmed communities property introduced in Section \ref{subsec:trimmed_communities}, all information required to explore the possible dynamic communities is contained within the set of active time nodes. This property asserts that L-Modularity is maximized when nodes enter a community at the time of their first interaction within it, and exit at the time of their last interaction. So extending a node's membership beyond its active participation leads to a reduction in the L-Modularity score. Considering time nodes with no interaction is then unnecessary.

LAGO starts by assigning each active time node to its own time module.


\subsection{Exploration step}

Greedy agglomerative approaches on static graphs rely on a local decision process for each node to update communities. Leveraging the sparsity of most real networks, one computes node by node the profit/loss in the global objective function of switching the node from its current community to the community of each of its neighbors. 

In LAGO, we need to adapt this approach to extend communities both topologically at a given time, and temporally.
We thus define the set of candidate time modules for changing attribution as the topological neighbors and the temporally adjacent time modules.

In the context of link streams, reassigning an active time node $ut$ from one time module to another involves two consequences:  (i) the time segment between $ut$ to its former module is no longer part of that module, and (ii) the time segment between $ut$ and its new module now belongs to the new module.

\subsubsection{Redefining the node movement operation} \label{sec:RTMM}

Contrary to the previously mentioned quality function used on static networks, L-Modularity is not \textit{linear} in the sense defined by \cite{campigotto2014generalizedadaptivemethodcommunity}. This implies that the impact of each local move cannot be precomputed independently of its context; it depends on the structure of the time modules involved in the move.
This non-linearity arises from the expectation term, which is not linear in time: the sum of ratios is not equal to the ratio of the sums. Nevertheless, the quality function is \textit{separable}, still in the sense of \cite{campigotto2014generalizedadaptivemethodcommunity}, meaning that the impact of a local move can be computed locally, without requiring a full recomputation of the global quality function.

As a consequence, (i) the graph aggregation phase used on static networks algorithms is not generalizable to link streams: while nodes may still be aggregated, time segments of varying durations within a time module cannot; and (ii) reassigning a sub-time module between time modules requires an evaluation of the profit/loss that involves all active time nodes involved in both the source and target time modules. This increases the computation cost of each move evaluation compared to what is done on static networks, as used in Louvain or Infomap. This underscores the importance of using fast exploration heuristic, as discussed in Section \ref{subsec:altap}.

\begin{figure}[h] 
\centering 
\subfloat[Exploration]{\includegraphics[width=.49\linewidth]{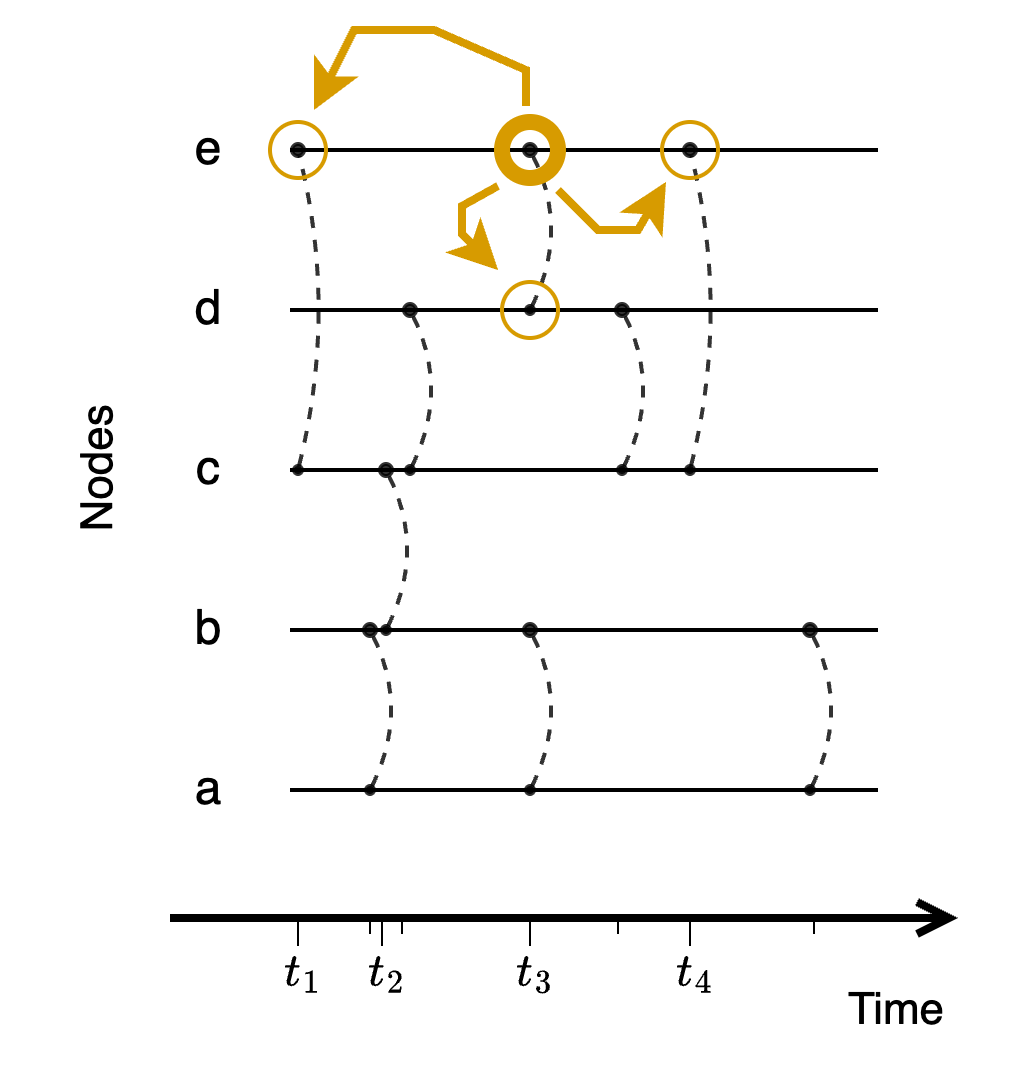}\label{figure:lada:a}}
\subfloat[Merging Time Modules]{\includegraphics[width=.49\linewidth]{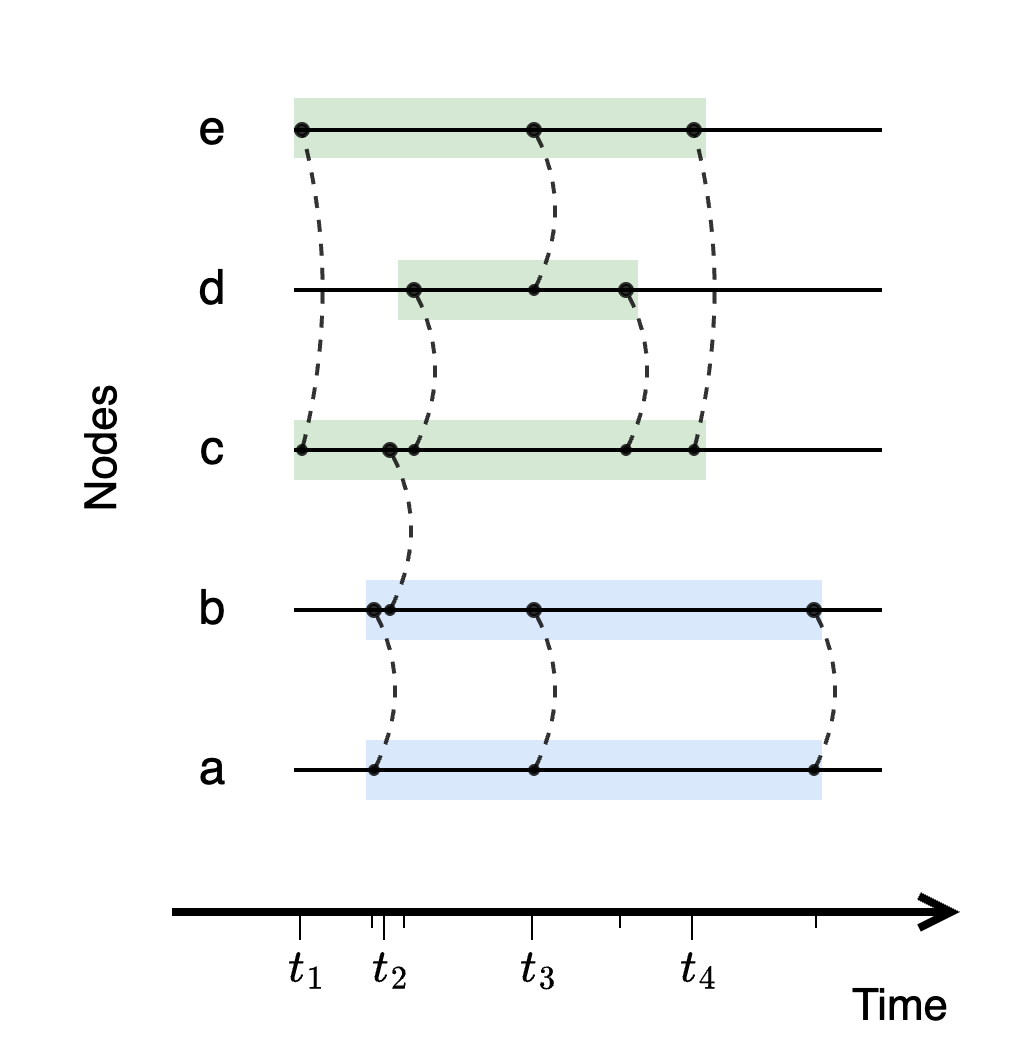}
\label{figure:lada:b}}
\caption{Illustration of the start of the Time Module Mover at the finest time modules level, where each active time node is first assigned to its own time module. 
Each active time node (e.g. $et_3$), is affiliated to the community that increases L-Modularity the most. Candidates for affiliation changes are topological neighbors (e.g. $dt_3$) and temporally adjacent time nodes (e.g. $et_1$, $et_4$). This process is repeated until no further affiliation change improves L-Modularity, leading to two time modules represented by the two colors} \label{figure:lada}
\end{figure}

To sum up, the core algorithm, called the \textit{Recursive Time Module Mover (RTMM)}, starts by assigning each active time node to its own time module. At any aggregation level of the algorithm, each time module is evaluated for merging with one of its neighbors: the topological neighbors and the time-adjacent modules. The move that improves the L-Modularity score the most is chosen. Note that each move trial requires (i) a computation based on all active time nodes from source and target time modules, and (ii) to compute the impact of the change of the time segments between the time modules. Figure \ref{figure:lada} illustrates the first phase of the algorithm at the finest aggregation level.

\subsection{Refinement steps}\label{subsec:refinement}

As is the case for static network exploration with Louvain, only applying the core algorithm phase can lead to convergence to a poor local optimum. It is caused by the greedy merging of time modules without the ability to revisit and refine previously formed structures. We expect this issue to be further exacerbated by the addition of the temporal setting, where the optimization process must account for both topological and temporal dimensions.
On static networks, refinement strategies have been proposed to overcome this limitation by enabling the reconsideration of prior decisions. We propose to adapt two such strategies, originally developed for Infomap \ref{subsec:infomap}, to the setting of link streams: (i) single-node movements, and (ii) submodule movements.

We also introduce the \textit{Single Time Edge Movement} refinement, a modification of the single-node movements tailored for link streams.

\subsubsection{Single Time Node Movements (STNM)}\label{sec:stnm}
Direct generalization of the single-node movement introduced by \cite{Rosvall_2009} consisting in moving single active time nodes between high level aggregated time modules if it increases L-Modularity.

\subsubsection{Sub Time Modules Movements (STMM)}\label{sec:stmm}
Direct generalization of the submodules movement introduced by \cite{Rosvall_2009} consisting in recomputing local submodules with RTMM solely on the induced sub link stream, and allowing them to change time module affiliation if it increases L-Modularity.

\subsubsection{Single Time Edge Movements (STEM)}\label{sec:stem}
This strategy generalizes Single Time Node Movements by allowing two interacting time nodes in the same community to be moved simultaneously. This joint movement addresses an expected limitation of independent node movements along the time axis: when attempting to reassign two connected active time nodes sequentially, the first move may result in a temporarily decrease in L-Modularity. This occurs because moving time nodes one by one temporarily extends the presence in a time module without contributing additional internal interaction, thereby increasing the expected number of interaction without a corresponding topological gain.
By contrast, moving the two interacting time nodes together preserves their topological contribution to the community and lead to an L-Modularity increase. To ensure that the benefits of individual time nodes exploration are not lost, the STEM strategy also incorporates the STNM strategy.


\subsection{Substitutions} \label{subsec:altap}

Other approaches have been proposed to improve greedy agglomerative optimization algorithms. In this section, we present two such substitutions, both incorporated into LAGO, which modify the application of either the core algorithm or its refinement phases.

\subsubsection{Fast Exploration (FE)}\label{sec:fe}

In the first phase of the Time Module Mover, iterating over all candidates time module and computing potential moves to neighbors implies a lot of computation with no guarantee of success. Moreover, the process starts again whenever at least one move in the list improves the quality function. As discussed in Section \ref{sec:RTMM}, evaluating the impact of a move requires recomputing the contribution of all active time nodes in both the source and target time modules.
In order to reduce the number of unnecessary computations we propose to adapt the efficient exploration strategy used in the Leiden algorithm\cite{Traag_2019}. We call it the Fast Exploration (FE). It is a heuristic that reduces the number of candidate neighbors to evaluate. The principle is to maintain a limited set of candidates, updated at each time module affiliation change. For each successful move, the neighbors of the moved time modules are added to the set of candidates. At each step, a candidate is selected and removed from the set for evaluation. The process continues until the set is empty, enabling localized but adaptive exploration of the solution space, while significantly reducing unnecessary computations.

\subsubsection{Refinement In RTMM (RIR)}\label{sec:rir}

In the context of static networks, refinement strategies are applied either after the main optimization loop \cite{Rosvall_2009} or within it \cite{Traag_2019}. Intuitively, applying refinement post hoc tends to result in faster algorithms, while incorporating it during the main loop may yield more fine-grained community structures. In this paper, we propose to compare the performance of both approaches when adapted to the context of link streams.

\begin{table}[h]
\begin{center}
\caption{LAGO variants. We consider 14 greedy algorithms, each representing a different combination of the steps introduced in this section. While all variants share the same foundational method --RTMM (Section \ref{sec:RTMM})--they differ in the specific refinements (Section \ref{subsec:refinement}) and substitutions (Section \ref{subsec:altap}) applied.} \label{tab1}
\begin{tabular}{|c|c|c|c|c|c|c|c|}
\hline
\textbf{}&\textbf{Core}&\multicolumn{3}{|c|}{\textbf{Refinement}}&\multicolumn{2}{|c|}{\textbf{Substitution}} \\
\cline{2-7} 
\textbf{Methods} & \textbf{RTMM}& \textbf{\textit{STNM}}& \textbf{\textit{STMM}}& \textbf{\textit{STEM}} & \textbf{\textit{FE}}& \textbf{\textit{RIR}} \\
\textbf{} & \ref{sec:RTMM}& \ref{sec:stnm}& \ref{sec:stmm}& \ref{sec:stem} & \ref{sec:fe} &  \ref{sec:rir} \\
\hline
$LV$ & v &  &  &  &  &   \\
$LV\star$ & v &  &  &  & v &   \\
$IM + N$ & v & v & v &  &  &   \\
$IM + N \star$ & v & v & v &  & v &   \\
$IM + E$ & v &   & v & v &  &   \\
$IM + E \star$ & v &  & v  & v & v  &   \\
$LV \times N$ & v & v &  &  &  &  v \\
$LV \times N \star$ & v & v &  &  & v & v  \\
$LV \times E$ & v &  &  & v &  & v  \\
$LV \times E\star$ & v &  &  & v & v & v  \\
$LV + N$ & v & v &  &  &  &   \\
$LV + N \star$ & v & v &  &  & v &   \\
$LV + E$ & v &  &  & v &  &   \\
$LV + E\star$ & v &  &  & v & v &   \\
\hline
\end{tabular}
\end{center}
\end{table}

\section{Experiments}\label{sec:experiments}

In this section, we demonstrate that LAGO is successful for optimizing L-Modularity, enabling the discovery of meaningful dynamic communities in link streams. We evaluate the different variants of LAGO (Table \ref{tab1}) with respect to scalability, optimization performance, and their ability to recover ground truth communities. Experiments are conducted on both synthetic and real-world temporal networks. Notably, a direct comparison with existing approaches is not possible, as no prior method has been proposed for the optimization of L-Modularity. Moreover, as highlighted by the authors of L-Modularity \cite{brabant2025longitudinal}, no publicly available method currently supports community detection directly on link streams. Our evaluation therefore focuses on comparing the performance of the different LAGO variants.

\subsection{Synthetic Dataset, Custom Community Structure}\label{subsec:sdccs}

\begin{figure*}[h]
  \centering
  \begin{minipage}[c]{0.18\textwidth}
    \centering
    \begin{minipage}[c]{1\textwidth} 
        \subfloat[Ground truth structure]{\includegraphics[width=\linewidth]{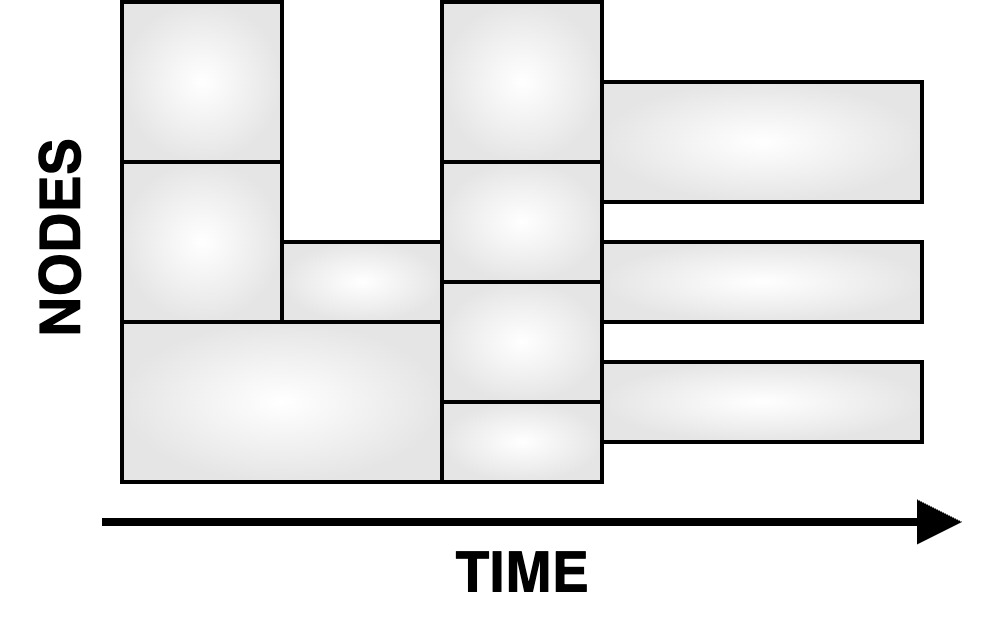}\label{figure:naive:gtcs}}
    \end{minipage}
    \hfill
    \begin{minipage}[c]{.95\textwidth}
        \subfloat[LAGO variants]{\includegraphics[width=\linewidth]{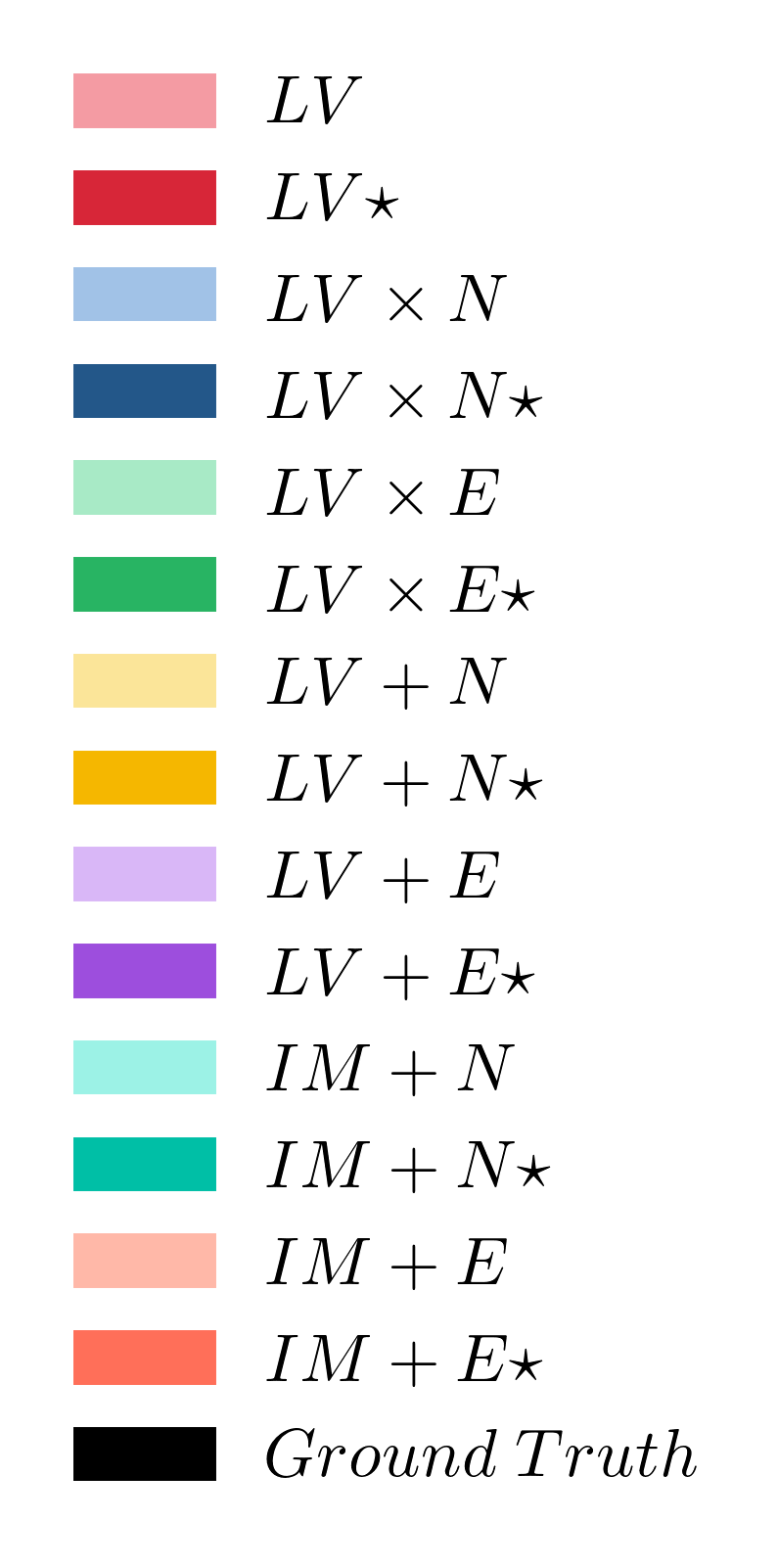}}
    \end{minipage}
  \end{minipage}
    \begin{minipage}[c]{0.81\textwidth} 
      \begin{minipage}[t]{1\textwidth}
        \centering
            \subfloat[Times of executions for optimizing $Q_{JM}$]{\includegraphics[width=.49\linewidth]{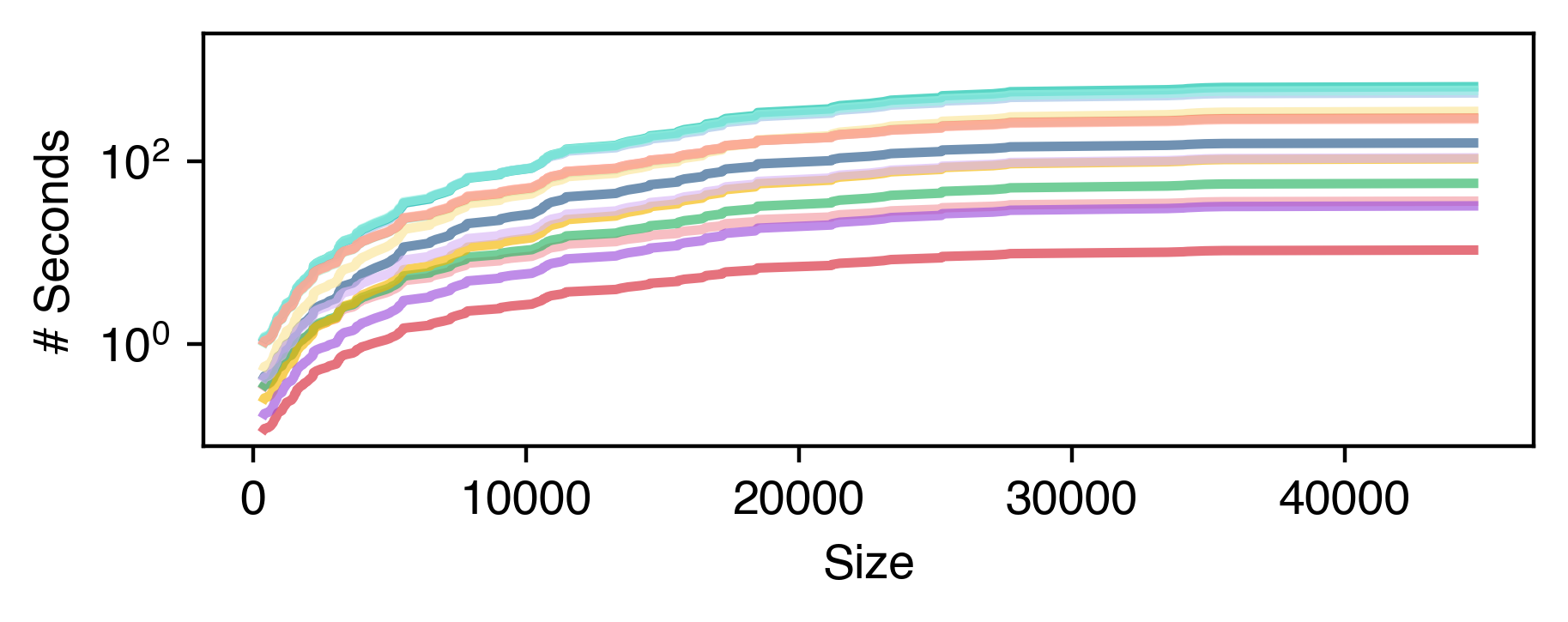}}
            \subfloat[Times of executions for optimizing $Q_{MM}$]{\includegraphics[width=.49\linewidth]{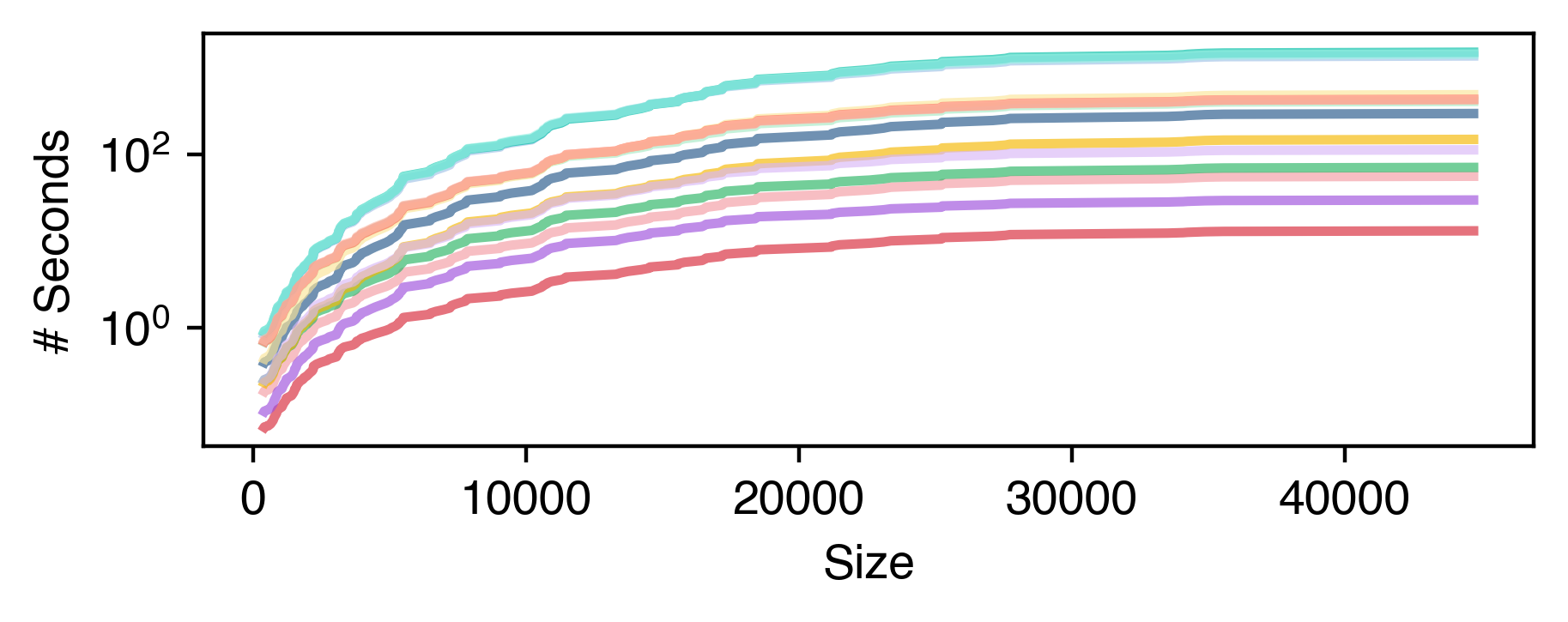}}
        \end{minipage}
        \begin{minipage}[t]{1\textwidth} 
        \centering
            \subfloat[$Q_{JM}$ values after optimization]{\includegraphics[width=.49\linewidth]{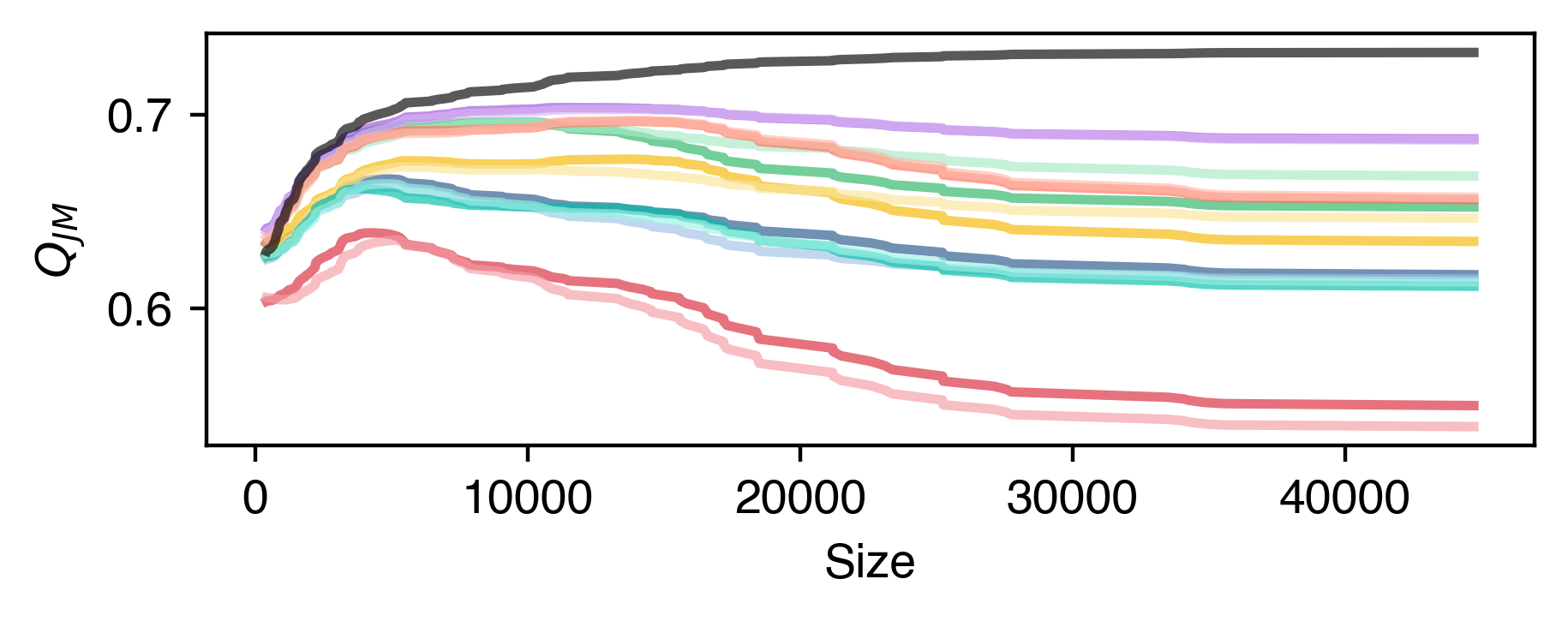}}
             \subfloat[$Q_{MM}$ values after optimization]{\includegraphics[width=.49\linewidth]{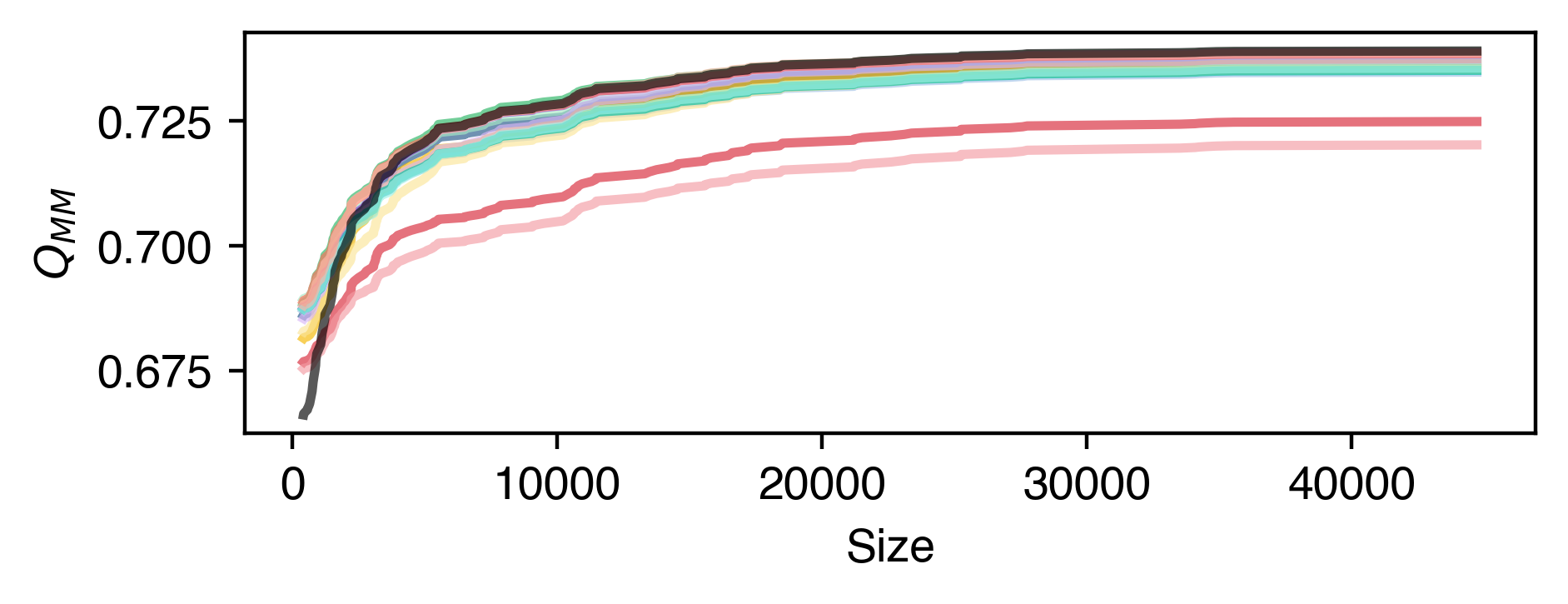}}
        \end{minipage}
         \begin{minipage}[t]{1\textwidth} 
         \centering
            \subfloat[$Q_{JM}$ optimization for ground truth communities retrieval]{\includegraphics[width=.49\linewidth]{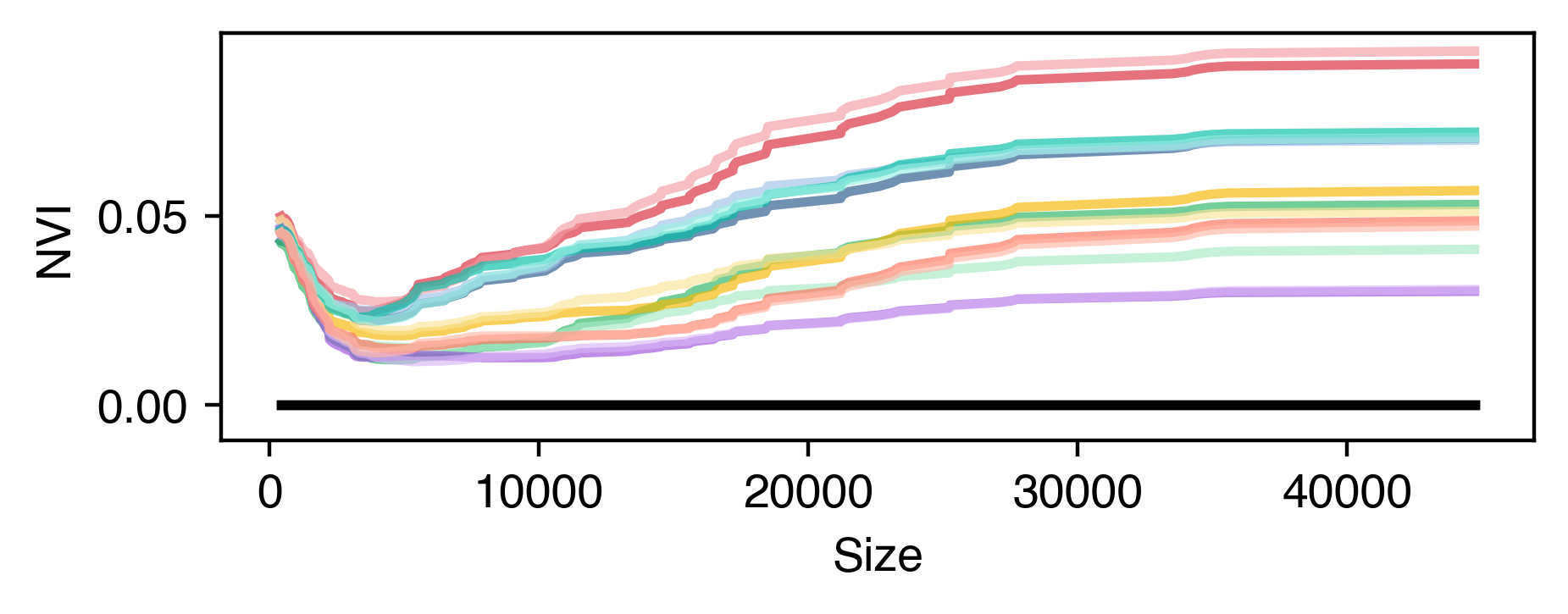}}
            \subfloat[$Q_{MM}$ optimization for ground truth communities retrieval]{\includegraphics[width=.49\linewidth]{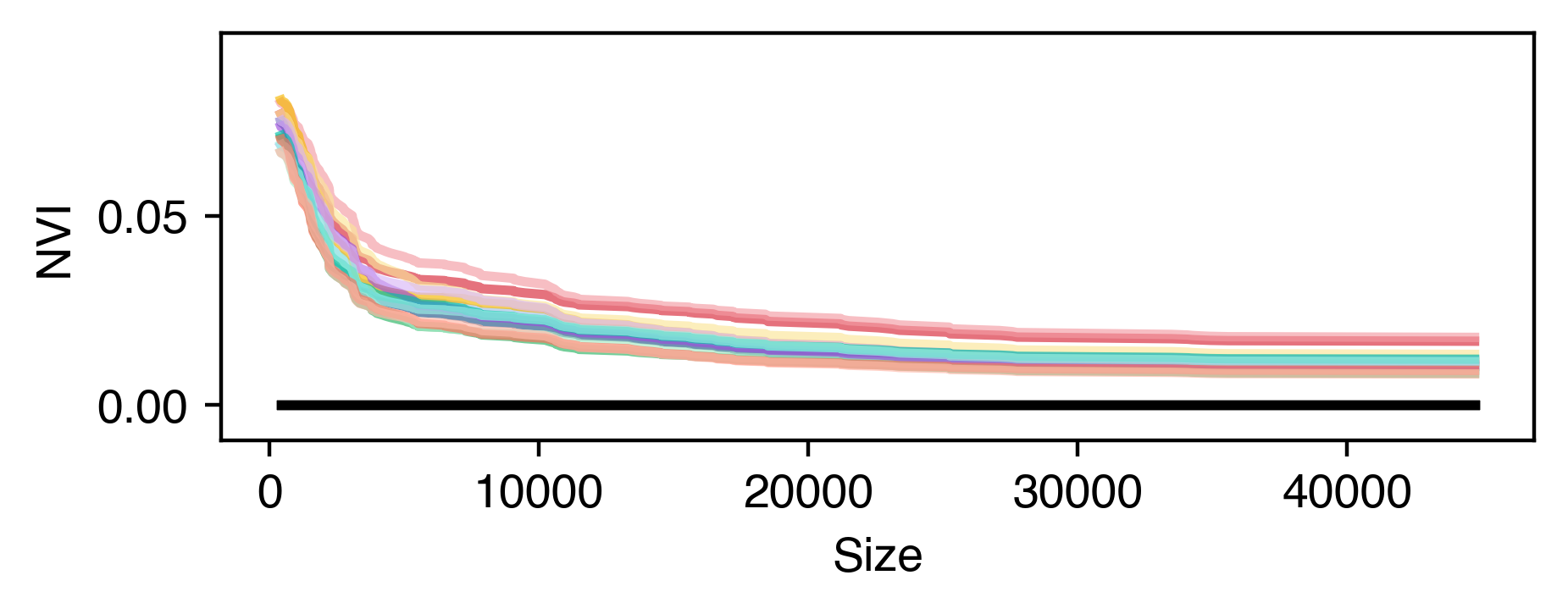}}
          \end{minipage}
      \end{minipage}

  \caption{Performance of LAGO variants in recovering the ground truth community structure (Fig.~\ref{figure:naive:gtcs}) across different link stream sizes. Both versions of L-Modularity are optimized. The following metrics are reported: 
  time of execution, final L-Modularity values, and the accuracy of community recovery as measured by the Normalized Variation of Information (NVI) between the detected and ground truth communities. The size of a link stream is the number of its active time nodes.}\label{figure:xp1}
\end{figure*} 

The first experiment demonstrates that LAGO is successful in recovering unambiguously defined dynamic communities. It also assesses its scalability.

To conduct this evaluation, we use the benchmark framework introduced by Asgari et al. \cite{asgari2023mosaicbenchmarknetworksmodular}, which enables the generation of link streams according to tailored ground truth community structures.

We first define a simple dynamic community structure illustrated in Fig.~\ref{figure:naive:gtcs}. Link streams are then generated by varying both the number of nodes and the maximum number of timesteps, proportionally scaling the community structure. Temporal edges are generated using the parameters from the benchmark framework, with an internal density coefficient $\alpha = 0.8$, and a community identifiability coefficient $\beta = 0$, indicating the absence of interactions between nodes belonging to different communities. 

We argue that an effective method for dynamic community detection should be capable of recovering these simple and well-defined dynamic communities. 

Figure \ref{figure:xp1} shows that LAGO successfully recovers the ground truth communities. This is evidenced by the high L-Modularity values---closely matching those of the ground truth---and the low Normalized Variation of Information (NVI) scores. Interestingly, optimizing $Q_{MM}$ yields better performance than optimizing $Q_{JM}$, despite the ground truth structure being more naturally aligned with the assumptions underlying $Q_{JM}$. We hypothesize that this is due to the more restrictive nature of $Q_{JM}$, which may lead the optimization to become trapped in local optima. 

Regardless of the LAGO variant used to optimize $Q_{JM}$, the results suggest that beyond a certain link stream size, the optimization process tends to a plateau, likely becoming trapped in suboptimal configurations. In contrast, optimization of $Q_{MM}$ appears more stable, achieving consistent performance across different link stream sizes.

Overall, the best-performing LAGO variant in this experiment, for both versions of L-Modularity, is $LV + E \star$. It consistently ranks among the top in terms of final L-Modularity values while also being the second fastest in execution time.

\subsection{Synthetic Dataset, Random Partitions}\label{subsec:sdrp}

The exploration of the space of possible temporal communities is influenced by numerous factors that remain largely underexplored. These include the number of nodes, the number of time steps, their ratio, the density of temporal interactions and their variations, as well as the choice of quality function, its parameters, and the associated optimization strategy. Furthermore, there is currently no universally accepted definition of what constitutes a ``good" dynamic community. To prevent evaluation bias, we avoid relying exclusively on arbitrarily predefined community structures when evaluating LAGO.

To address these limitations, we evaluate LAGO's performance across a broad spectrum of synthetic link streams and dynamic community structures, randomly generated. We adopt a relative performance approach, comparing the 14 LAGO variants on a diverse set of generated link streams to identify performance trends. For each instance, the variants are ranked independently on two criteria: (i) execution time, and (ii) the final value of the optimized L-Modularity. Rankings are assigned from best (rank 1) to worst (rank 14) per criterion.

The synthetic link streams are generated using the framework proposed by Asgari et al. \cite{asgari2023mosaicbenchmarknetworksmodular}. Parameters are sampled uniformly: the number of nodes between 25 and 250; the maximum number of timesteps between 50 and 250; the internal density coefficient $\alpha$ between 0.5 and 1; and the inter-community interaction coefficient $\beta$ between 0 and 0.5, subject to the constraint $\beta < \alpha / 3$ to ensure clearly separable communities. Each LAGO variant is executed three times per link stream, and only the median values are retained for evaluation. It enables more reliable results since the greedy methods are based on randomness.
The average size of the generated link streams is 10,122 active time nodes, with a standard deviation of 8,468.

\begin{figure}[h]
\centering
\subfloat[Performances distributions of LAGO optimizing $Q_{JM}$]{\includegraphics[width=.9\linewidth]{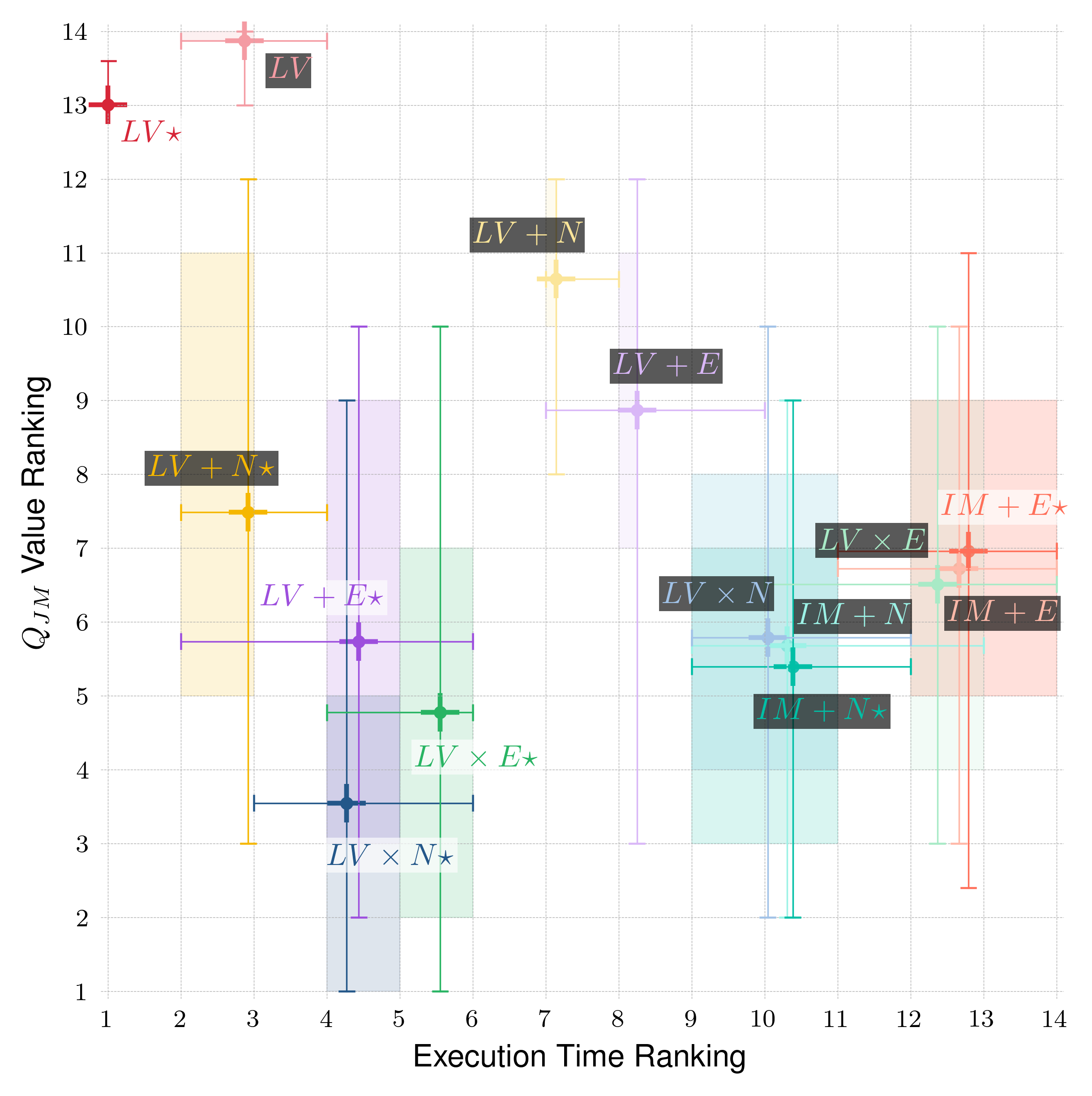}
\label{fig_first_case}}
\hfil
\subfloat[Performances distributions of LAGO optimizing $Q_{MM}$]{\includegraphics[width=.9\linewidth]{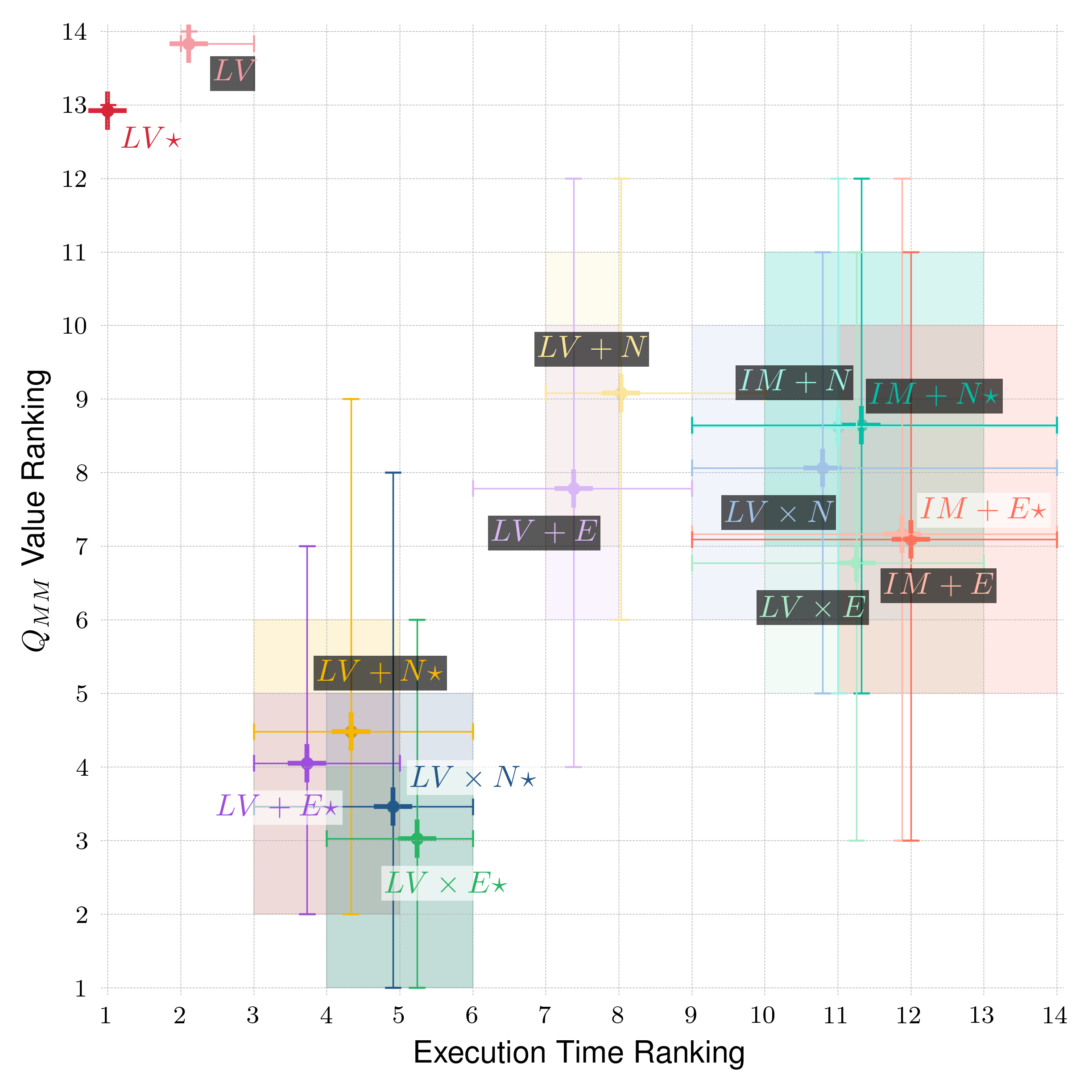}
\label{fig_second_case}}
\caption{Relative comparisons of LAGO variants (see Table \ref{tab1}) when optimizing both L-Modularity versions. For a given variant, the cross indicates the mean ranking of time execution and the mean L-Modularity value ranking, areas are for the values between the 1st and 3rd quartile of each axe, and the whiskers indicates values between the 1st and 9th deciles. Best LAGO variants are in the bottom left corner.}
\label{figure:onrandom}
\end{figure}

Figure~\ref{figure:onrandom} presents the comparative results of the LAGO variants. We can observe a Pareto front of optimality composed of very few methods, 
with those in the bottom left dominating most others.
For both L-Modularity objectives, $LV$ and $LV \star$ consistently achieve the fastest execution times but yield the lowest L-Modularity scores. This highlights the importance of incorporating a refinement phase in the algorithm, as these two variants are the only ones with no such incorporated step.


Furthermore, variants without the Fast Exploration option tend to be slower, with the notable exception of $IM+N \star$ and $IM+E \star$, both of which incorporate STMM. STMM is computationally expensive due to the non-linearity \cite{campigotto2014generalizedadaptivemethodcommunity} of L-Modularity and requires considering all active time nodes involved in a move---of both source and target module---to compute the related improvement.

Interestingly, incorporating Fast Exploration does not degrade optimization performance. On the contrary, LAGO variants that include this strategy generally outperform those that do not. This supports the hypothesis that prioritizing the neighbors of previously moved time modules is a more effective exploration strategy than random selection. 

Additionally, LAGO variants that incorporate the "Refinement In RTMM"
mechanism generally exhibit longer execution times compared to their
counterparts without this mechanism.
This is consistent with the increased number of loop iterations and candidate evaluations required by the refinement process.

Among LAGO variants, $LV \times N \star$ achieves the best results for optimizing $Q_{JM}$, while $LV \times E \star$ performs best for $Q_{MM}$. Nevertheless, both variants demonstrate strong performance across both objective functions. The inclusion of the STEM mechanism appears particularly advantageous for $Q_{MM}$, while STNM is more effective for $Q_{JM}$. This may be explained by the fact that in $Q_{JM}$, even a slight extension of a node’s membership duration in a community directly influences the community’s total duration and thus its contribution to the L-Modularity score.

In conclusion, this experiment reveals that more exhaustive move trials do not necessarily lead to better optimization outcomes; on the contrary, excessive exploration may increase the risk of becoming trapped in poor local optima. The STMM mechanism appears to be significantly less effective and could be excluded from future configurations. All effective variants include Fast Exploration, underscoring its practical value. The final choice between LAGO flavors should depend on specific use-case requirements.

\subsection{Real Data Application}\label{subsec:rda}

\begin{figure*}[h]
\centering
\subfloat[Dynamic communities obtained with LAGO]{\includegraphics[width=1\linewidth]{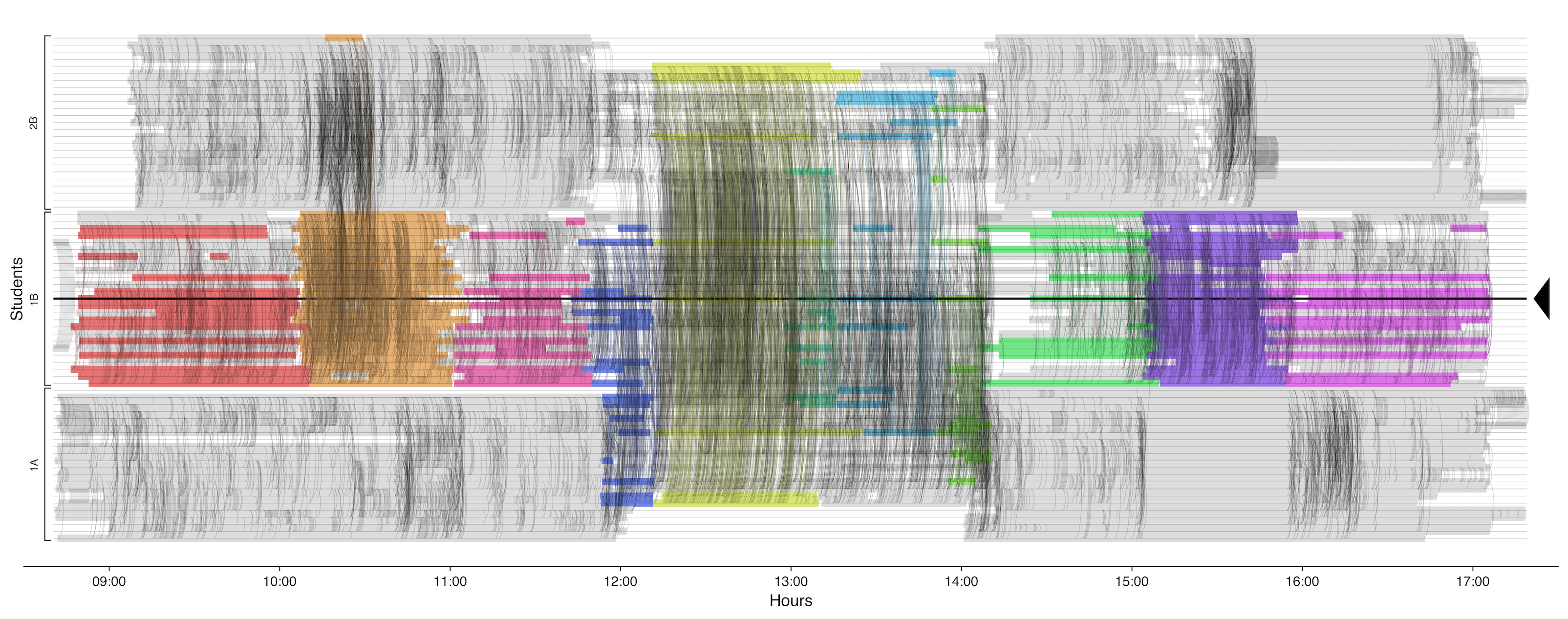}
\label{figure:socio:lago}}
\hfil
\subfloat[Dynamic communities obtained by Multi Slice Modularity optimization]{\includegraphics[width=1\linewidth]{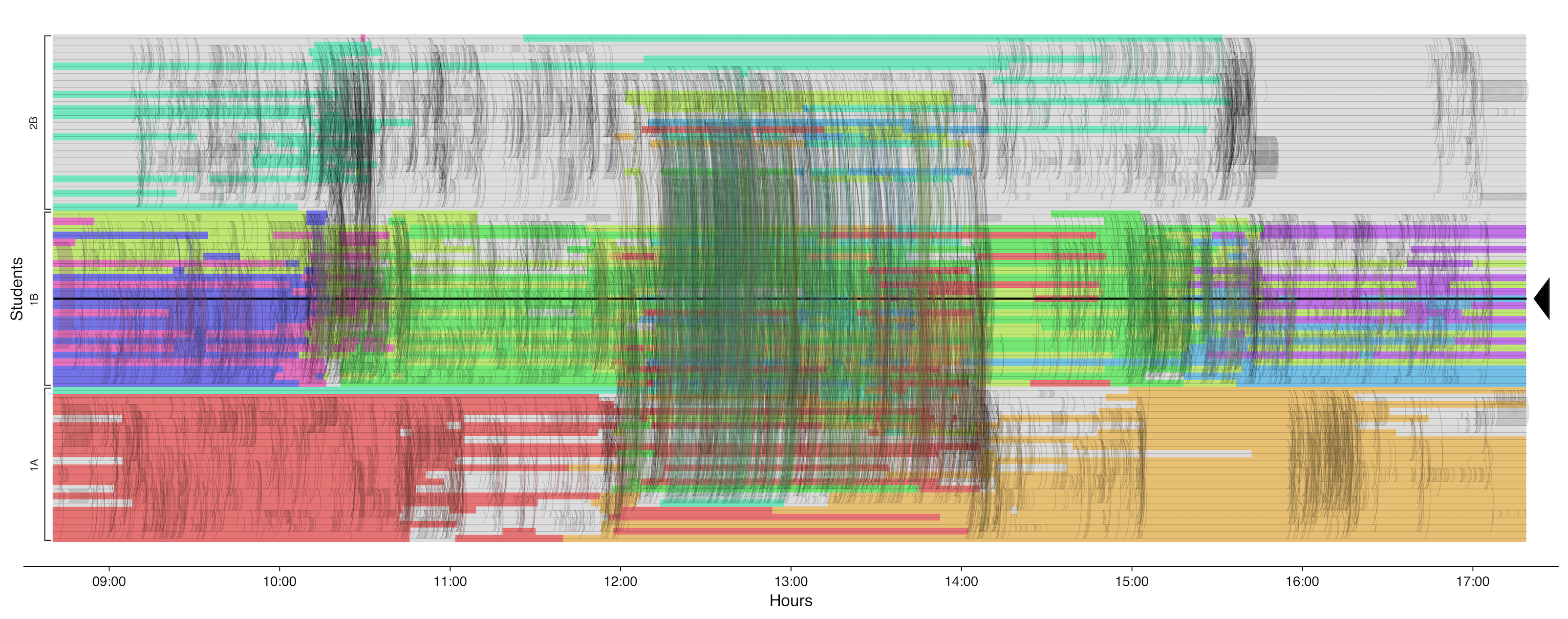}
\label{figure:socio:mucha}}
\caption{Two dynamic community structures identified from temporal interactions in the Primary School SocioPatterns dataset, restricted to three classes on the first day. Students are represented as horizontal lines; vertical curved lines indicate moments of face-to-face interaction. Colors show the communities visited by the student highlighted in bold (indicated by the arrowhead on the right); all other communities are shown in grey.}
\label{figure:socio}
\end{figure*}

In this section, we evaluate LAGO on real-world temporal data and compare its performance with the closest existing method for dynamic community detection based on modularity: multislice modularity \cite{Mucha_2010}, an extension of modularity designed to detect dynamic communities in temporal networks represented as sequences of static networks.

We use the primary school socio-pattern dataset \cite{10.1371/journal.pone.0023176}, which records interactions between pairs of students every 20 seconds when they face each other. The dataset also includes class membership information for each student. It allows for observing the dynamics of interactions over few days with different phases alternating between lectures and breaks.

In the absence of reliable ground truth, performance is assessed qualitatively through visualization of the link stream and detected dynamic communities. Due to limitations imposed by the size of the dataset, both in terms of the number of nodes and the number of time steps, we focus on interactions among three classes during the first day of recording. This subset contains 72 students, 1,555 time steps, and 28,904 active time nodes.

We apply the $LV\times N \star$ variant of LAGO to optimize $Q_{JM}$, with a temporal smoothing parameter $\omega=15$ on the link stream. We compare it with multislice modularity, using an existing implementation \footnote{https://github.com/vtraag/leidenalg}. We chose the weights parameters in order to produce a manageable number of communities for interpretation. 

The results are shown in Figure \ref{figure:socio}. The visualization focuses on a single student, highlighted with a bold horizontal black line. Colored communities are those in which the student belongs during at least one time step. LAGO identifies 11 dynamic communities for this student. 
The first three communities visited by the student---colored red, orange, and pink---primarily consist of classmates, though with varying patterns of affiliation. Each community lasts approximately one hour, matching the duration of a lecture in that school. Similarly, the last three communities (green, purple, and magenta) reflect different interactions with students from the same class. During the lunch break, the student visits five communities, each comprising different subsets of students from all the classes. 

In contrast, the communities detected by the multislice modularity method (Figure~\ref{figure:socio:mucha}) appear less meaningful.  This is unsurprising, as the multislice approach was not designed to handle rapid interactions and frequent changes characteristic of this dataset.

It is worth noting that the multislice approach treats the link stream as a sequence of snapshots, requiring consideration of $|V| \times |T| = 72 \times 1555 = 111 960$ time nodes. By contrast, LAGO only considers $28 904$ active time nodes, leading to more efficient computation. 

\subsection{Results Highlights}

Experiments demonstrate that LAGO effectively optimizes L-Modularity, though its efficiency varies with the choice of variant. Performance depends on both the target objective---$Q_{JM}$ or $Q_{MM}$---and the structure of the link stream. This is illustrated by the first two experiments, ranging from simple (Section~\ref{subsec:sdccs}) to complex, randomly generated link streams (Section~\ref{subsec:sdrp}), and do not consistently favor the same LAGO variants for optimal performance. Nevertheless, the experiments suggest general guidelines for effective LAGO use:
\begin{itemize}
\item \textbf{Refinement strategy:} Strongly recommended. While STNM generally yields better performance for $Q_{JM}$ and STEM for $Q_{MM}$, this association is not absolute and may vary depending on the context. The use of STMM is not recommended.
\item \textbf{Fast Exploration:} Strongly recommended. This mechanism significantly reduces computation time and improves L-Modularity scores.
\item \textbf{Refinement in RTMM:} Optional. It tends to improve results but increases computational cost.
\end{itemize}

\section{Conclusion}

This paper introduced and evaluated LAGO, a greedy optimization algorithm for Longitudinal Modularity thereby establishing the first method designed to detect communities directly in continuous-time networks, without requiring a predefined time scale for analysis. 

Through a comprehensive comparative analysis of 14 LAGO variants, we assessed the influence of design choices. These results reveal general trends in performance and suggest guidelines for selecting appropriate strategies depending on optimization goals. Our findings also point to the importance of further investigating the structural properties of temporal networks and their communities.

Finally, the LAGO framework is compatible for optimizing other quality functions defined over link streams and sharing a similar design with the L-Modularity.

\section*{Code Availability}
All code necessary to reproduce the experiments is available at \url{https://osf.io/cqtnj/file}. An open-source Python library implementing the most effective LAGO variants is provided at \url{https://github.com/fondationsahar/dynamic_community_detection}.

\section*{Acknowledgment}
We thank SAHAR for financing this project, and also Yasaman Asgari from the University of Zurich for her precious feedbacks during the writing of this paper.

\bibliographystyle{unsrt}
\bibliography{references}

\begin{thebibliography}{10}

\bibitem{Blondel_2008}
Vincent~D Blondel, Jean-Loup Guillaume, Renaud Lambiotte, and Etienne Lefebvre.
\newblock Fast unfolding of communities in large networks.
\newblock {\em Journal of Statistical Mechanics: Theory and Experiment}, 2008(10):P10008, October 2008.

\bibitem{Rosvall_2009}
M.~Rosvall, D.~Axelsson, and C.~T. Bergstrom.
\newblock The map equation.
\newblock {\em The European Physical Journal Special Topics}, 178(1):13–23, November 2009.

\bibitem{Traag_2019}
V.~A. Traag, L.~Waltman, and N.~J. van Eck.
\newblock From louvain to leiden: guaranteeing well-connected communities.
\newblock {\em Scientific Reports}, 9(1), March 2019.

\bibitem{latapy2018stream}
Matthieu Latapy, Tiphaine Viard, and Cl{\'e}mence Magnien.
\newblock Stream graphs and link streams for the modeling of interactions over time.
\newblock {\em Social Network Analysis and Mining}, 8:1--29, 2018.

\bibitem{rossetti2018community}
Giulio Rossetti and R{\'e}my Cazabet.
\newblock Community discovery in dynamic networks: a survey.
\newblock {\em ACM computing surveys (CSUR)}, 51(2):1--37, 2018.

\bibitem{Mucha_2010}
Peter~J. Mucha, Thomas Richardson, Kevin Macon, Mason~A. Porter, and Jukka-Pekka Onnela.
\newblock Community structure in time-dependent, multiscale, and multiplex networks.
\newblock {\em Science}, 328(5980):876–878, May 2010.

\bibitem{infomap_multilayer}
Ulf Aslak, Martin Rosvall, and Sune Lehmann.
\newblock Constrained information flows in temporal networks reveal intermittent communities.
\newblock {\em Phys. Rev. E}, 97:062312, Jun 2018.

\bibitem{matias_2015}
Catherine Matias and Vincent Miele.
\newblock Statistical clustering of temporal networks through a dynamic stochastic block model.
\newblock {\em Journal of the Royal Statistical Society: Series B (Statistical Methodology)}, 79, 06 2015.

\bibitem{bovet2022flow}
Alexandre Bovet, Jean-Charles Delvenne, and Renaud Lambiotte.
\newblock Flow stability for dynamic community detection.
\newblock {\em Science advances}, 8(19):eabj3063, 2022.

\bibitem{brabant2025longitudinal}
Victor Brabant, Yasaman Asgari, Pierre Borgnat, Angela Bonifati, and R{\'e}my Cazabet.
\newblock Longitudinal modularity, a modularity for link streams.
\newblock {\em EPJ Data Science}, 14(1):12, 2025.

\bibitem{newman2004finding}
Mark~EJ Newman and Michelle Girvan.
\newblock Finding and evaluating community structure in networks.
\newblock {\em Physical review E}, 69(2):026113, 2004.

\bibitem{PhysRevE.84.016114}
V.~A. Traag, P.~Van~Dooren, and Y.~Nesterov.
\newblock Narrow scope for resolution-limit-free community detection.
\newblock {\em Phys. Rev. E}, 84:016114, Jul 2011.

\bibitem{campigotto2014generalizedadaptivemethodcommunity}
Romain Campigotto, Patricia~Conde Céspedes, and Jean-Loup Guillaume.
\newblock A generalized and adaptive method for community detection, 2014.

\bibitem{asgari2023mosaicbenchmarknetworksmodular}
Yasaman Asgari, Remy Cazabet, and Pierre Borgnat.
\newblock Mosaic benchmark networks: Modular link streams for testing dynamic community detection algorithms, 2023.

\bibitem{10.1371/journal.pone.0023176}
Juliette Stehlé, Nicolas Voirin, Alain Barrat, Ciro Cattuto, Lorenzo Isella, {Jean-François} Pinton, Marco Quaggiotto, Wouter {Van den Broeck}, Corinne Régis, Bruno Lina, and Philippe Vanhems.
\newblock High-resolution measurements of face-to-face contact patterns in a primary school.
\newblock {\em PLOS ONE}, 6(8):e23176, 08 2011.

\end{thebibliography}

\end{document}